# Artificial Neural Network Based Prediction of Optimal Pseudo-Damping and Meta-Damping in Oscillatory Fractional Order Dynamical Systems

Saptarshi Das[1], Indranil Pan[1,2]

1. Department of Power Engineering, Jadavpur University, Salt-Lake Campus, LB-8, Sector 3, Kolkata-700098, India. Email: saptarshi@pe.jusl.ac.in
2. MERG, Energy, Environment, Modelling and Minerals ($E^2M^2$) Research Section, Department of Earth Science and Engineering, Imperial College London, Exhibition Road, London SW7 2AZ, UK. Email: i.pan11@imperial.ac.uk, indranil.jj@student.iitd.ac.in

Khrist Sur[1,3], Shantanu Das[4]

3. Center for Soft Computing Research, Indian Statistical Institute, 203 Barrackpore Trunk Road, Kolkata-700108, India. Email: khrist@ieee.org, khrist@isical.ac.in
4. Reactor Control Division, Bhabha Atomic Research Centre, Mumbai-400085, India.
Email: shantanu@magnum.barc.gov.in

*Abstract*—This paper investigates typical behaviors like damped oscillations in fractional order (FO) dynamical systems. Such response occurs due to the presence of, what is conceived as, pseudo-damping and meta-damping in some special class of FO systems. Here, approximation of such damped oscillation in FO systems with the conventional notion of integer order damping and time constant has been carried out using Genetic Algorithm (GA). Next, a multilayer feed-forward Artificial Neural Network (ANN) has been trained using the GA based results to predict the optimal pseudo and meta-damping from knowledge of the maximum order or number of terms in the FO dynamical system.

## I. INTRODUCTION

Fractional order dynamical systems which are governed by fractional order differential equations have got renewed interest in the science and engineering community in recent past for its higher capability and flexibility in modeling of natural processes [1], [2]. It is well known from basics of control theory that second order stable oscillatory dynamical systems or higher order stable oscillatory systems, which can also be approximated as second order transfer function models, decay with an exponential envelope upon step or impulse type excitation [3]. In other words, a physical system, governed by second order differential equation of the form (1), with excitation $u(t)$ and response $y(t)$ shows oscillatory time response for $\xi \in (0,1)$.

$$\frac{d^2 y(t)}{dt^2} + 2\xi\omega\frac{dy(t)}{dt} + \omega^2 y(t) = \omega^2 u(t)$$

$$\Rightarrow \tau^2 \frac{d^2 y(t)}{dt^2} + 2\xi\tau\frac{dy(t)}{dt} + y(t) = u(t) \quad (1)$$

Here, parameters $\{\xi, \omega, \tau\}$ represent the system's damping ratio, natural frequency and time constant respectively with $\tau = 1/\omega$. For step and impulse type excitation, the dynamical system governed by (1) exhibit damped time responses with an exponential envelope, represented by (2) and (3) respectively.

$$y(t) = 1 - \frac{e^{-\xi\omega t}}{\sqrt{1-\xi^2}} \sin\left(\omega t\sqrt{1-\xi^2} + \tan^{-1}\left(\frac{\sqrt{1-\xi^2}}{\xi}\right)\right) \quad (2)$$

$$y(t) = \frac{\omega e^{-\xi\omega t}}{\sqrt{1-\xi^2}} \sin\left(\omega t\sqrt{1-\xi^2}\right) \quad (3)$$

In contrast, a dynamical system, governed by a two term fractional order differential equation (4) can also show oscillatory damped time response for $\alpha \in (1, 2)$, although there is no explicit damping term, containing $\xi$ in (4). This typical behavior of fractional order systems lead to the concept of "Pseudo-damping" which can not be visualized with the conventional theory of integer order calculus for describing the dynamics of physical systems.

$$aD_t^\alpha y(t) + by(t) = u(t) \quad (4)$$

Laplace transform of (4) with zero initial condition gives the system's transfer function as (5) which again produces its impulse response as the Green's function (6) upon inverse Laplace transformation [1] for the two-term FO system (4).

$$G_2(s) = \frac{Y(s)}{U(s)} = \frac{1}{as^\alpha + b} = \frac{1}{a}\left(\frac{1}{s^\alpha + (b/a)}\right) \quad (5)$$

$$g_2(t) = \frac{1}{a}t^{\alpha-1}E_{\alpha,\alpha}\left(-\frac{b}{a}t^\alpha\right) \quad (6)$$

In (6), $E_{\alpha,\beta}$ represents the two-parameter Mittag-Leffler function which is a higher transcendental, encompassing a large family of conventional transcendental functions like trigonometric, inverse circular, exponential, logarithmic, hyperbolic etc. [1]-[2]. The series representation of two parameter Mittag-Leffler function is given by (7) which is a generalized template and reduces to an exponential function for $\alpha = 1, \beta = 1$ [1]-[2]. Also, from (6) it can be observed that the envelope is guided by a power law instead of an exponential one in (3). In this system, a Mittag-Leffler type oscillation takes place in contrast to the sinusoidal oscillation in (3).

$$E_{\alpha,\beta}(z) := \sum_{k=0}^{\infty} \frac{z^k}{\Gamma(\alpha k + \beta)}, \quad \alpha > 0, \beta > 0 \quad (7)$$

The present paper firstly attempts to approximate the oscillations, produced due to step excitation of (4) with an equivalent template given by (1) using GA i.e. finding optimal pseudo/meta (FO)-damping or time constants, associated with the oscillatory time response of the FO system. Next the optimal FO damping and time constants are predicted using a multilayer feed-forward ANN. This approach reduces the computational load, associated with running GA every time for finding out the equivalent optimal FO damping for any arbitrary FO system within this range and such an application is justified from the point that multilayer feed-forward ANN is generally very good function approximator [4]-[5]. In [6], the concept of optimal fractional order damping was first proposed with a specific need for faster stabilization of oscillatory systems using the concept of FO damping with respect to some integral performance indices like Integral of Squared Error (ISE), Integral Time weighted Squared Error (ITSE) etc. This paper gives a new concept of finding the optimal integer order equivalence of FO damping using ISE/ITSE as performance indices and also proposes their ANN based prediction.

Rest of the paper is organized as follows. Section II briefly introduces the basics of pseudo-damping and meta-damping in some special class of oscillatory FO dynamical system. Section III describes time domain simulation of pseudo/meta-damping and their GA based optimal time domain approximation. Section IV presents the ANN based training and prediction performance for these optimal FO-damping. The paper ends with the conclusion as section V, followed by the references.

## II. CONCEPT OF PSEUDO AND META-DAMPING IN FRACTIONAL ORDER LINEAR DYNAMICAL SYSTEMS

In [1]-[2], it has been reported that time and frequency domain representations of few special functions, related to fractional calculus like R-function and G-function are given by:

$$\mathcal{L}^{-1}\left[\frac{s^v}{s^\alpha - a}\right] = \sum_{n=0}^{\infty} \frac{a^n t^{(n+1)\alpha - 1 - v}}{\Gamma((n+1)\alpha - v)} \quad (8)$$

$$\mathcal{L}^{-1}\left[\frac{s^v}{(s^\alpha - a)^r}\right] = \sum_{j=0}^{\infty} \frac{\{(-r)(-1-r)\cdots(1-j-r)\}(-a)^j t^{(r+j)\alpha - v - 1}}{\Gamma(1+j)\Gamma((r+j)\alpha - v)} \quad (9)$$

If Laplace transform of excitation and response of a fractional order transfer function (FOTF) $G(s)$ be $U(s)$ and $Y(s)$ respectively, then simple treatments as in (10) gives its step and impulse response. Therefore, to find out the step and impulse response for few classes of fractional order systems, $v = -1$ and $v = 0$ need to be considered respectively in (8) and (9). Time domain simulation for FO systems using (8)-(9) needs evaluation of few convergent infinite series at each discrete time step ($t$). For numerical implementation the infinite series have been evaluated in MATLAB at each $t$, with the order of accuracy being 0.001.

$$\begin{aligned} Y(s) &= G(s)U(s) \\ \Rightarrow y(t) &= g(t) * u(t) \\ &= \begin{cases} \mathcal{L}^{-1}[G(s)/s] & \text{for step response} \\ \mathcal{L}^{-1}[G(s)] & \text{for impulse response} \end{cases} \end{aligned} \quad (10)$$

Under this condition, expressions (8) and (9) represents the step response of stable FOTFs with the replacement of ($a$) by unity and ($b$) by ($-b$) in structures like (4). i.e.

$$P_1 = \frac{1}{s^\alpha + b}, \quad P_2 = \frac{1}{(s^\alpha + b)^r} \quad (11)$$

It has been seen from (4) that the step response gives sustained oscillation for $\alpha = 0$. But the response becomes damped for $1 < \alpha < 2$. Hence, a fractional order system of the form (4) having no explicit damping term in it, also exhibits damped oscillation in time response for $1 < \alpha < 2$. Such an oscillation has been approximated by using a second order system of the form (1) while minimizing few integral error indices. The FO system of the structure (11) with $1 < \alpha < 2$, can be modified with normalized frequency to unity ($b = 1$) as:

$$\tilde{P}_1 = \frac{1}{s^\alpha + 1} \simeq \frac{1}{\tau^2 s^2 + 2\tau\xi s + 1} \quad (12)$$

where, $\{\tau, \xi\}$ can be termed as the optimal pseudo-time constant and pseudo-damping respectively with respect to some integral error index.

The second class of FO systems in (11) exhibits different type of oscillations if different combinations of orders are used in the expansion of the polynomials, although the highest order of the models are same and only the number of fractional order terms varies in the model. It is well known that order of a FO LTI system is determined by the maximum order present in the denominator polynomial. If it be assumed that $r = 2/\alpha$, the system governed by (9) becomes a different class of fractional second order system (13) which can again be represented by equivalent second order approximation with $\{\tau, \xi\}$ being the optimal meta-time constant and meta-damping respectively. Similar treatment of normalizing the frequency to unity yields:

$$\tilde{P}_2 = \frac{1}{(s^\alpha + 1)^{2/\alpha}} \simeq \frac{1}{\tau^2 s^2 + 2\tau\xi s + 1} \quad (13)$$

The following examples put more light on the behavior of such systems with meta-damping. Simple modification of (13) gives first and second order transfer functions like (14). It is interesting to note that though the leading order remains one and two in these models, the number of fractional order elements increase upon binomial expansion for the terms with higher powers. These additional number of FO terms puts extra damping to the FO system which is defined as the meta-damping in FO dynamical system of the form (13). Thus a FO system represented by (14) is distinctly characterized by the number of FO elements present in it and not by the leading FO order unlike (12). This typical behavior is the motivation behind defining two different class of FO damping i.e. pseudo-damping for system (12) and meta-damping for system (13).

$$\frac{1}{\left(s^\alpha+1\right)^{1/\alpha}} = \frac{1}{\left(s^{0.2}+1\right)^5}, \frac{1}{\left(s^{0.25}+1\right)^4}, \frac{1}{\left(s^{0.5}+1\right)^2}, \frac{1}{\left(s^{0.1}+1\right)^{10}}, \cdots$$
$$\frac{1}{\left(s^\alpha+1\right)^{2/\alpha}} = \frac{1}{\left(s^{0.2}+1\right)^{10}}, \frac{1}{\left(s^{0.25}+1\right)^8}, \frac{1}{\left(s^{0.4}+1\right)^5}, \frac{1}{\left(s^{0.5}+1\right)^4}, \cdots$$
(14)

It is therefore clear that the pseudo-damping is associated with the reduction in the highest order of a FO system whereas meta-damping is associated with the increase in fractional order elements within a FO model though the highest order of the plant remains the same.

### III. TIME DOMAIN SIMULATION OF FRACTIONAL ORDER SYSTEMS WITH PSEUDO-DAMPING AND META-DAMPING

MATLAB based codes have been developed using the infinite series representations of such FOTF i.e. (8)-(9) under impulse/step excitation. Time domain simulation using (8)-(9) often gives poor result above 30 seconds. This is due to the fact that gamma function in the denominator of (8)-(9) approaches towards a very large value which can not be computed using most of the scientific programming languages, due to buffer overflow. Thus it is recommended to reliably use expression (8)-(9) for time domain simulation of the special class of FO systems only up to 30 seconds. Simulation of first order system with meta-damping and $\alpha < 0.9$ also becomes computationally infeasible due to blowing up of the associated gamma functions. Similarly, second order systems with meta-damping and $\alpha < 0.9$ gives reliable time response up to 25 seconds, below which the results are unreliable as also reported in Hartley and Lorenzo [6] in the context of optimal FO damping.

*A. Step Response Characteristics*

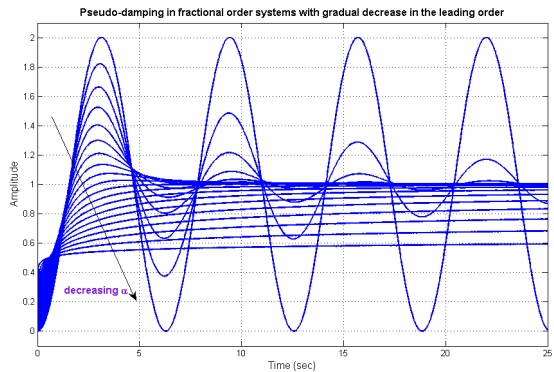

Figure 1. Step response of FO system with pseudo-damping.

FO systems given in (12)-(13) have now been subjected to step input excitations and evaluated at each discrete time step using (8)-(9) and shown in Fig. 1-3. Fig. 1 shows that the oscillations become more damped with decrease in the order ($\alpha$) of FO system (12). Similar behaviors can be found for FO systems with meta-damping with leading order being 2 (Fig. 2) and 1 (Fig. 3) respectively. It is interesting to note that even first order systems in the presence of other FO elements may exhibit oscillations as shown in Fig. 3. In Fig. 1 the decaying envelope may be guided by a power law as reported in (6), but the nature of oscillations for meta-damping in Fig. 2-3 are more complex to be represented as closed form solutions unlike (6).

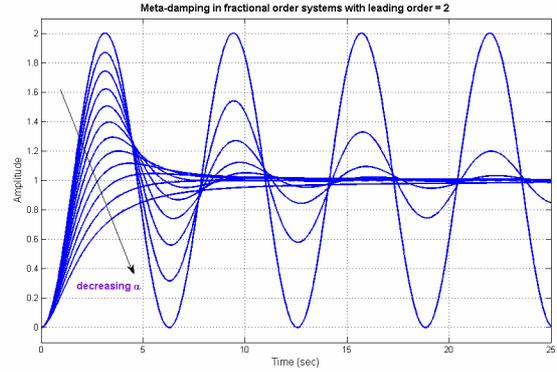

Figure 2. Step response characteristics of fractional second order system with meta-damping.

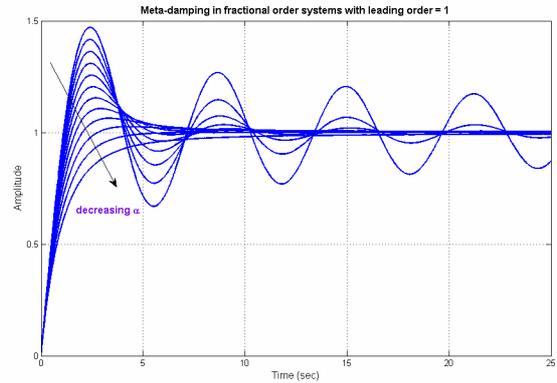

Figure 3. Step response characteristics of fractional first order system with meta-damping.

*B. Impulse Response Characteristics*

The impulse responses have been shown next for the above discussed three classes of FO systems. As expected in Fig. 6, the oscillations start from a value of unity for first order systems (with additional FO elements causing meta-damping). Fig. 5 shows the oscillations starting from zero confirming the preservation of the second order behavior of the FO system. In Fig. 4 showing FO systems with pseudo-damping mixed behaviors can be observed regarding the initial value of the impulse response which indicates that the classical notion of judging the order of the system by looking only at the impulse response characteristics is not valid for pseudo-damping.

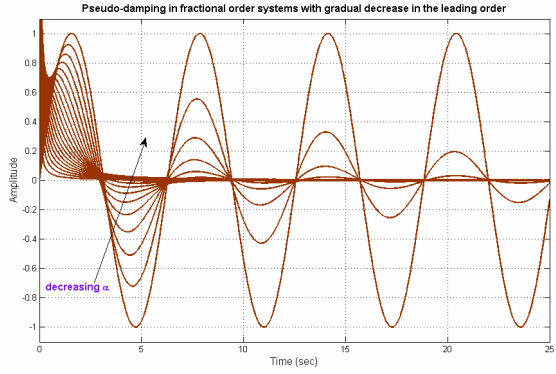

Figure 4. Impulse response of FO system with pseudo-damping.

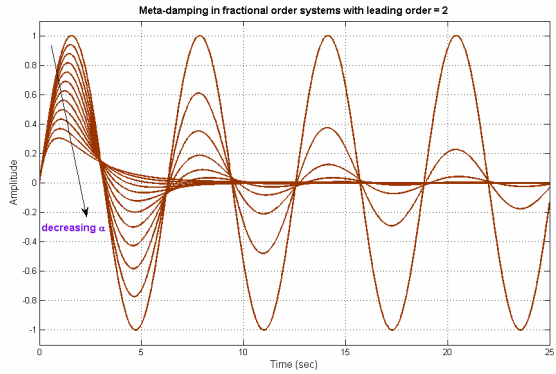

Figure 5. Impulse response characteristics of fractional second order system with.meta-damping.

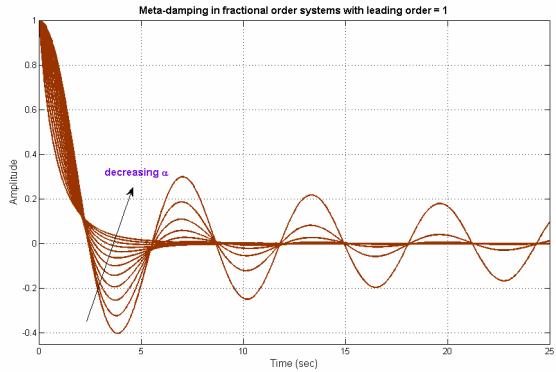

Figure 6. Impulse response characteristics of fractional first order system with meta-damping.

## C. Genetic Algorithm Based Approach for Finding Optimal Pseudo-Damping and Meta-Damping

Genetic algorithm (GA) is a stochastic optimization process which can be used to minimize a chosen objective function. A solution vector is initially randomly chosen from the search space and undergoes reproduction, crossover and mutation, in each iteration to give rise to a better population of solution vectors in the next iteration. Reproduction implies that solution vectors with higher fitness values can produce more copies of themselves in the next generation. Crossover refers to information exchange based on probabilistic decisions between solution vectors. In mutation a small randomly selected part of a solution vector is occasionally altered, with a very small probability. This way the solution is refined iteratively until the objective function is minimized below a certain tolerance level or the maximum number of iterations are exceeded. In the present study the number of population members in GA is chosen to be 20. The crossover and mutation fraction are chosen to be 0.8 and 0.2 respectively for minimization of the following objective functions.

$$J_{ISE} = \int_0^\infty e^2(t)dt, \quad J_{ITSE} = \int_0^\infty t \cdot e^2(t)dt \quad (15)$$

Evaluation of the objective functions have been done in each generation of GA within the finite time horizon of 25 seconds as discussed earlier. Minimization of error index (15) between the respective FO systems and a second order approximation as in (13) and (14) gives the optimum values of pseudo/meta-damping and time constant. The ISE and ITSE based optimization results have been reported in Table I-III for the three test cases, as in Fig. 1-3.

TABLE I. GA BASED RESULTS FOR OPTIMAL PSEUDO-DAMPING FOR FRACTIONAL ORDER SYSTEMS

| Fractional Order (α) | ISE Based | | | ITSE Based | | |
|---|---|---|---|---|---|---|
| | $J_{min}$ | $\tau$ | $\xi$ | $J_{min}$ | $\tau$ | $\xi$ |
| 1.1 | 0.0054 | 0.3485 | 1.3152 | 0.0235 | 0.47 | 0.9848 |
| 1.2 | 0.0168 | 0.5246 | 0.8094 | 0.0647 | 0.6467 | 0.6887 |
| 1.3 | 0.0287 | 0.6587 | 0.596 | 0.0979 | 0.7659 | 0.5537 |
| 1.4 | 0.0379 | 0.7634 | 0.4672 | 0.1153 | 0.8457 | 0.4635 |
| 1.5 | 0.0429 | 0.8432 | 0.374 | 0.1176 | 0.8998 | 0.3878 |
| 1.6 | 0.0429 | 0.9029 | 0.2965 | 0.1085 | 0.9374 | 0.3146 |
| 1.7 | 0.0378 | 0.9463 | 0.2247 | 0.0927 | 0.9647 | 0.239 |
| 1.8 | 0.0277 | 0.976 | 0.1527 | 0.074 | 0.9837 | 0.1597 |
| 1.9 | 0.0127 | 0.993 | 0.0771 | 0.0448 | 0.9947 | 0.0786 |

TABLE II. GA BASED RESULTS FOR OPTIMAL META-DAMPING FOR FRACTIONAL ORDER SYSTEMS WITH LEADING ORDER = 1

| Fractional Order (α) | ISE Based | | | ITSE Based | | |
|---|---|---|---|---|---|---|
| | $J_{min}$ | $\tau$ | $\xi$ | $J_{min}$ | $\tau$ | $\xi$ |
| 1.1 | 0.0057 | 0.3463 | 1.1946 | 0.02 | 0.4006 | 1.0538 |
| 1.2 | 0.0147 | 0.342 | 1.067 | 0.0512 | 0.4933 | 0.7747 |
| 1.3 | 0.0273 | 0.4182 | 0.7884 | 0.0769 | 0.5761 | 0.6247 |
| 1.4 | 0.0415 | 0.4793 | 0.6322 | 0.0956 | 0.6399 | 0.5336 |
| 1.5 | 0.0571 | 0.5317 | 0.5308 | 0.1105 | 0.6906 | 0.4667 |
| 1.6 | 0.0748 | 0.5996 | 0.4393 | 0.1271 | 0.7352 | 0.4091 |
| 1.7 | 0.097 | 0.6597 | 0.3708 | 0.1568 | 0.7784 | 0.3529 |
| 1.8 | 0.13 | 0.7233 | 0.3086 | 0.2382 | 0.8253 | 0.2913 |
| 1.9 | 0.1996 | 0.7959 | 0.2422 | 0.5806 | 0.8828 | 0.2131 |

TABLE III. GA BASED RESULTS FOR OPTIMAL META-DAMPING FOR FRACTIONAL ORDER SYSTEMS WITH LEADING ORDER = 2

| Fractional Order (α) | ISE Based | | | ITSE Based | | |
|---|---|---|---|---|---|---|
| | $J_{min}$ | $\tau$ | $\xi$ | $J_{min}$ | $\tau$ | $\xi$ |
| 1.1 | 0.005 | 1.0426 | 0.7299 | 0.0462 | 1.0837 | 0.712 |
| 1.2 | 0.0138 | 1.0499 | 0.574 | 0.1053 | 1.0939 | 0.5732 |
| 1.3 | 0.0215 | 1.0452 | 0.4668 | 0.1359 | 1.0809 | 0.4809 |
| 1.4 | 0.0266 | 1.0362 | 0.3845 | 0.1407 | 1.0609 | 0.4066 |
| 1.5 | 0.0291 | 1.0263 | 0.3153 | 0.1315 | 1.0409 | 0.3395 |
| 1.6 | 0.029 | 1.0171 | 0.2529 | 0.117 | 1.0241 | 0.2741 |
| 1.7 | 0.0266 | 1.0094 | 0.1931 | 0.1038 | 1.012 | 0.2083 |
| 1.8 | 0.0217 | 1.0042 | 0.1325 | 0.0935 | 1.0047 | 0.1405 |
| 1.9 | 0.0119 | 1.0012 | 0.0679 | 0.0657 | 1.0012 | 0.0701 |

## IV. ANN BASED PREDICTION OF OPTIMAL PSEUDO-DAMPING AND META-DAMPING

### A. Multi-Layer Feedforward Neural Network Architecture

The standard neural network architecture consists of an input layer, one or more hidden layers with multiple perceptrons and an output layer. The number of perceptrons in the hidden layer and the number of hidden layers are generally problem specific and depend on the choice of the user. In the present study, the number of hidden layers is varied from 1 to 2 and for each case the number of neurons in each layer is varied from 5 to 25 in incremental steps of 5. The ANN is fully connected, i.e., the output from each input and hidden neuron is distributed to all the neurons of the subsequent layer. Also a feed-forward architecture is used, i.e. the data flows and is processed sequentially through the input, hidden and output layers and are not fed-back to the previous layers, unlike the recurrent ANN structure. Hyperbolic tangent sigmoid (*tansig*) and logarithmic sigmoid (*logsig*) type activation functions and their combinations are used to create different ANN architectures for comparing the relative effectiveness of these structures at capturing the nonlinear relationship between the input ($\alpha$) and output ($\tau, \xi$) data.

### B. Training Performance of ANN and Time Domain Performance of the Predicted Outputs

Multilayer feed-forward ANN has now been employed to predict the optimal pseudo/meta damping/time constants from the knowledge of the fractional order of the dynamical system. Tables IV-VI gives the GA based optimal pseudo/meta damping ($\xi$) and time constant ($\tau$) of the FO systems in terms of equivalent second order systems considering the ITSE criterion, since ITSE puts more penalties on the error at later stages unlike ISE producing better accuracy. The ANNs are trained with Levenberg-Marquardt back-propagation algorithm which is a gradient based method and often gets stuck in local minima. To check the consistency of the ANNs for capturing the input-output relationship, the Mean Squared Error (MSE) of 25 independent runs has been chosen as the performance measure. This is justified from the fact that often large size multilayer ANNs accurately establishes arbitrary nonlinear relation between any input-output data but they might not show consistency in the mapping [4]-[5]. Hence there is always a trade-off between size of the ANN and the prediction accuracy.

TABLE IV. TRAINING PERFORMANCE FOR VARIOUS ANN CONFIGURATIONS FOR THE PREDICTION OF PSEUDO-DAMPING FOR 25 INDEPENDENT RUNS

| Number of layers | Number of neurons in each hidden layer | Activation function | Average MSE |
|---|---|---|---|
| 1 | 5 | *tansig* | 0.009 |
| | | *logsig* | 0.0061 |
| | 10 | *tansig* | 0.07 |
| | | *logsig* | 0.0326 |
| | 15 | *tansig* | 0.1009 |
| | | *logsig* | 0.026 |
| | 20 | *tansig* | 0.1609 |
| | | *logsig* | 0.0418 |
| | 25 | *tansig* | 0.1216 |
| | | *logsig* | 0.0559 |
| 2 | 5 | *tansig/tansig* | 0.0185 |
| | | *tansig/logsig* | 0.0148 |
| | | *logsig/tansig* | 0.0238 |
| | | *logsig/logsig* | 0.0153 |
| | 10 | *tansig/tansig* | 0.0453 |
| | | *tansig/logsig* | 0.0169 |
| | | *logsig/tansig* | 0.0218 |
| | | *logsig/logsig* | 0.0133 |
| | 15 | *tansig/tansig* | 0.0714 |
| | | *tansig/logsig* | 0.0245 |
| | | *logsig/tansig* | 0.0779 |
| | | *logsig/logsig* | 0.0346 |
| | 20 | *tansig/tansig* | 0.1075 |
| | | *tansig/logsig* | 0.0269 |
| | | *logsig/tansig* | 0.1492 |
| | | *logsig/logsig* | 0.0375 |
| | 25 | *tansig/tansig* | 0.141 |
| | | *tansig/logsig* | 0.0341 |
| | | *logsig/tansig* | 0.2229 |
| | | *logsig/logsig* | 0.0357 |

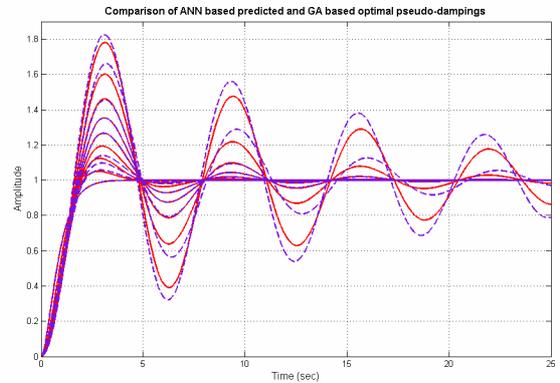

Figure 7. ANN prediction of pseudo-damping for FO system (12).

TABLE V. TRAINING PERFORMANCE FOR VARIOUS ANN CONFIGURATIONS FOR THE PREDICTION OF META-DAMPING IN FO SYSTEMS WITH LEADING ORDER = 1 FOR 25 INDEPENDENT RUNS

| Number of layers | Number of neurons in each hidden layer | Activation function | Average MSE |
|---|---|---|---|
| 1 | 5 | tansig | 0.0109 |
| | | logsig | 0.0038 |
| | 10 | tansig | 0.0487 |
| | | logsig | 0.0232 |
| | 15 | tansig | 0.0942 |
| | | logsig | 0.0257 |
| | 20 | tansig | 0.1113 |
| | | logsig | 0.026 |
| | 25 | tansig | 0.1002 |
| | | logsig | 0.0694 |
| 2 | 5 | tansig/tansig | 0.0126 |
| | | tansig/logsig | 0.0117 |
| | | logsig/tansig | 0.0086 |
| | | logsig/logsig | 0.0074 |
| | 10 | tansig/tansig | 0.0385 |
| | | tansig/logsig | 0.021 |
| | | logsig/tansig | 0.0392 |
| | | logsig/logsig | 0.0121 |
| | 15 | tansig/tansig | 0.0763 |
| | | tansig/logsig | 0.0257 |
| | | logsig/tansig | 0.0414 |
| | | logsig/logsig | 0.0314 |
| | 20 | tansig/tansig | 0.0985 |
| | | tansig/logsig | 0.0201 |
| | | logsig/tansig | 0.0934 |
| | | logsig/logsig | 0.0596 |
| | 25 | tansig/tansig | 0.0991 |
| | | tansig/logsig | 0.0303 |
| | | logsig/tansig | 0.0973 |
| | | logsig/logsig | 0.0281 |

TABLE VI. TRAINING PERFORMANCE FOR VARIOUS ANN CONFIGURATIONS FOR THE PREDICTION OF META-DAMPING IN FO SYSTEMS WITH LEADING ORDER = 2 FOR 25 INDEPENDENT RUNS

| Number of layers | Number of neurons in each hidden layer | Activation function | Average MSE |
|---|---|---|---|
| 1 | 5 | tansig | 0.0032 |
| | | logsig | 0.0014 |
| | 10 | tansig | 0.0274 |
| | | logsig | 0.0091 |
| | 15 | tansig | 0.0686 |
| | | logsig | 0.0093 |
| | 20 | tansig | 0.0431 |
| | | logsig | 0.019 |
| | 25 | tansig | 0.0654 |
| | | logsig | 0.0436 |
| 2 | 5 | tansig/tansig | 0.0058 |
| | | tansig/logsig | 0.0059 |
| | | logsig/tansig | 0.0038 |
| | | logsig/logsig | 0.0043 |
| | 10 | tansig/tansig | 0.014 |
| | | tansig/logsig | 0.0099 |
| | | logsig/tansig | 0.0126 |
| | | logsig/logsig | 0.0057 |
| | 15 | tansig/tansig | 0.0451 |
| | | tansig/logsig | 0.006 |
| | | logsig/tansig | 0.0353 |
| | | logsig/logsig | 0.0154 |
| | 20 | tansig/tansig | 0.0451 |
| | | tansig/logsig | 0.0304 |
| | | logsig/tansig | 0.0339 |
| | | logsig/logsig | 0.0139 |
| | 25 | tansig/tansig | 0.0372 |
| | | tansig/logsig | 0.0137 |
| | | logsig/tansig | 0.0351 |
| | | logsig/logsig | 0.0206 |

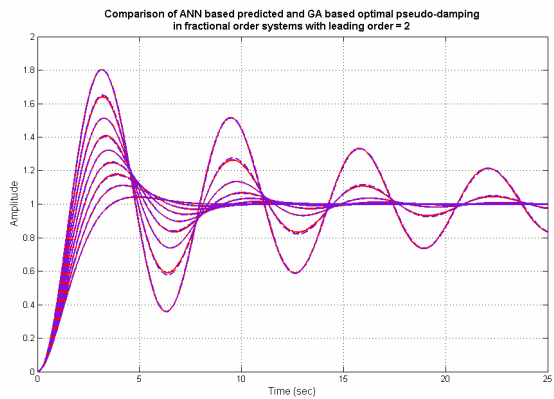

Figure 8. ANN based prediction of meta-damping for fractional second order system.

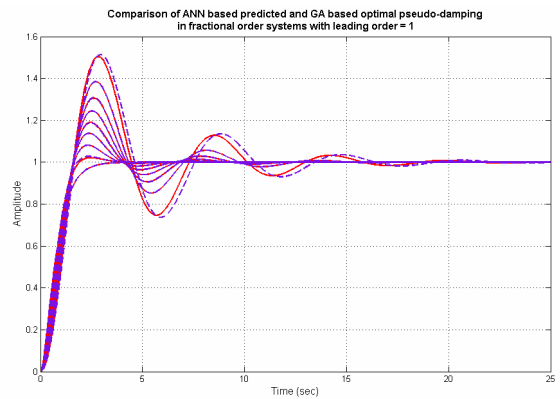

Figure 9. ANN based prediction of meta-damping for fractional first order system.

TABLE VII. ANN BASED PREDICTED VALUES OF PSEUDO/META-DAMPINGS AND TIME CONSTANTS FOR THREE CLASS OF FO SYSTEMS

| Type of the FO System | Minimum MSE | Fractional Order (α) | τ | ξ |
|---|---|---|---|---|
| with pseudo-damping | $6.1938 \times 10^{-4}$ | 1.1 | 0.469965 | 0.984699 |
| | | 1.2 | 0.69043 | 0.673145 |
| | | 1.3 | 0.760275 | 0.595913 |
| | | 1.4 | 0.819033 | 0.531421 |
| | | 1.5 | 0.898188 | 0.387535 |
| | | 1.6 | 0.937233 | 0.314444 |
| | | 1.7 | 0.966328 | 0.242967 |
| | | 1.8 | 1.009048 | 0.130724 |
| | | 1.9 | 0.986992 | 0.061456 |
| First order with meta-damping | $9.6422 \times 10^{-5}$ | 1.1 | 0.400597 | 1.053798 |
| | | 1.2 | 0.495297 | 0.751861 |
| | | 1.3 | 0.576102 | 0.624702 |
| | | 1.4 | 0.639898 | 0.5336 |
| | | 1.5 | 0.688546 | 0.46988 |
| | | 1.6 | 0.7352 | 0.4091 |
| | | 1.7 | 0.779042 | 0.350727 |
| | | 1.8 | 0.8253 | 0.2913 |
| | | 1.9 | 0.916861 | 0.207589 |
| Second order with meta-damping | $1.4428 \times 10^{-5}$ | 1.1 | 1.091173 | 0.707394 |
| | | 1.2 | 1.088813 | 0.573404 |
| | | 1.3 | 1.076325 | 0.4878 |
| | | 1.4 | 1.062211 | 0.402436 |
| | | 1.5 | 1.040513 | 0.339679 |
| | | 1.6 | 1.019453 | 0.278533 |
| | | 1.7 | 1.012378 | 0.208497 |
| | | 1.8 | 1.006116 | 0.13578 |
| | | 1.9 | 1.002982 | 0.070353 |

In IV-VI the ANN with 5 neurons in the single hidden layer with *logsig* activation function has been found to produce consistently good prediction of the optimal FO damping in terms of minimum average MSE for 25 runs. Subsequent increase in number of layers or different combinations of activation functions may result in lower MSE in some particular cases. But considering the consistency of ANNs, these complicated structures have been found to produce always a higher value of average MSE. Table VII reports the predicted values and MSE of three different classes of FO systems for different fractional orders while considering single hidden layer containing 5 neurons and *logsig* activation function. The simulations and results justify the argument that complicated ANN topology may establish arbitrary mapping but the consistency of such mapping reduces with large size of the network [4]-[5]. Time response curves for the systems with the predicted pseudo-meta damping and time constants corresponding to that presented Table VII and the GA based optimum results in Tables I-III has been shown in Fig. 7-9 for step input excitation. In Fig 7-9 it is clear that the ANN based predicted results (dashed lines) are very close to the GA based optimum values (continuous lines). The prediction is very accurate for first and second order meta-damping than the pseudo-damping and also for low value of fractional order (α).

In contemporary literatures like [7]-[9], the concept of pseudo-damping was first proposed with a different second order like structure of the FO systems with commensurable orders. In the present paper, pseudo-damping refers to the damping introduced in the system with decrease in its leading order which is different from that reported in [6]-[9]. The contribution of the present paper is firstly to give systematic definition of pseudo and meta-damping in FO systems and secondly their ANN based prediction from the GA based ITSE optimum results.

## V. CONCLUSION

In this paper, the concepts of pseudo-damping and meta-damping are introduced for some special class of fractional order dynamical systems. Genetic algorithm is used to obtain the equivalent second order damping characteristics of these FO systems. An ANN based approach is used next to model this arbitrary (nonlinear) relationship and eliminate the requirement of running the computationally intensive GA every time. Extensive parametric study have been done to find out the multilayer feed-forward ANN architecture that is capable to optimally capture the nonlinearity, by minimizing the MSE and simultaneously avoiding the pitfall of over-fitting. Consistency of ANN structures for mapping the fractional order to optimal FO damping and time constants are judged by considering average MSE of 25 independent runs. The time domain comparison of the GA based optimum results with the ANN based predicted results is found to be very close.